\documentclass[3p,number]{elsarticle}

\usepackage{amsmath}
\usepackage{amssymb}
\usepackage{graphicx}
\usepackage{color}

\usepackage{times}

\begin{document}

\begin{frontmatter}

\title{Stability criteria for complex ecosystems \tnoteref{CONT}} 
\tnotetext[CONT]{The Authors contributed equally.}

\journal{Science}

\author[eee,ci]{Stefano Allesina} 
\ead{sallesina@uchicago.edu}

\author[eee]{Si Tang}
\ead{sugar@uchicago.edu}

\address[eee]{Dept. Ecology \& Evolution, University of Chicago, 1101
  E. 57th Chicago, IL 60637 USA.}

\address[ci]{Computation Institute, University of Chicago.}

\begin{abstract}
  Forty years ago, Robert May questioned a central belief in ecology
  by proving that sufficiently large or complex ecological networks
  have probability of persisting close to zero. To prove this point,
  he analyzed large networks in which species interact at
  random. However, in natural systems pairs of species have
  well-defined interactions (e.g., predator-prey, mutualistic or
  competitive). Here we extend May's results to these relationships
  and find remarkable differences between predator-prey interactions,
  which increase stability, and mutualistic and competitive, which are
  destabilizing. We provide analytic stability criteria for all
  cases. These results have broad applicability in ecology. For
  example, we show that, surprisingly, the probability of stability
  for predator-prey networks is decreased when we impose realistic
  food web structure or we introduce a large preponderance of weak
  interactions. Similarly, stability is negatively impacted by
  nestedness in bipartite mutualistic networks.
\end{abstract}

\end{frontmatter}

May showed mathematically that large and complex ecosystems are
inherently unstable \citep{may1972will,may2001stability}. This
contribution has been one of the main drivers of theoretical ecology
ever since
\citep{pimm1984complexity,mccann2000diversity,montoya2006ecological},
as it clashed with the prevailing belief of ecologists that large,
highly complex ecosystems (such as those observed empirically) were
more stable than simpler ones (found in extreme environments and
disturbed ecosystems)
\citep{macarthur1955fluctuations,elton2001animal}.

May's theorem deals with a particular type of community matrix
\citep{levins1968evolution,may1972will,may2001stability} $M$, of size
$S\times S$ ($S$ is the number of species in the system). The matrix
$M$ describes the effect a species $j$ (column) has on species $i$
(row) around the equilibrium point of an unspecified dynamical system
describing the density of the species through time.

The diagonal coefficients of $M$ are all $-1$, while the off-diagonal
coefficients are drawn from a normal distribution $N(0,\sigma^2)$ with
probability $C$ and are $0$ otherwise. For large $S$, May proved that
the probability of stability is close to $0$ whenever the
``complexity'' $K=\sigma \sqrt{SC}>1$. Local stability measures the
tendency of the system to return to equilibrium after small
perturbations. In unstable systems, even infinitesimal perturbations
will make the system move away from the equilibrium state, potentially
resulting in the loss of species. Thus, it should be extremely
improbable to observe rich (large $S$) or highly connected (large $C$)
ecosystems persisting through time. Mathematically, an equilibrium
point is stable if all the eigenvalues of the corresponding community
matrix have negative real part.

The networks described by these matrices have random structure: each
pair of species interacts with a given probability. However, this
randomness translates, for large $S$, into fixed interaction
frequencies, so that when we are constructing the matrices above, we
are following a precise mixture of interaction types. We can classify
interactions types according to the signs of the ordered pair
$(M_{ij},M_{ji})$ (effect of $j$ on $i$ and vice versa), and compute
their expected frequencies in a large random matrix. For the pair
$(0,0)$: non-interacting, the expected frequency is $(1-C)^2$; $(+,0)$
or $(0,+)$: commensalism, $C(1-C)$; $(-,0)$ or $(0,-)$: amensalism,
$C(1-C)$; $(-,-)$: competition, $C^2/4$; $(+,+)$: mutualism, $C^2/4$;
$(+,-)$ or $(-,+)$: predator-prey, $C^2/2$.

Here we show how the criterion for stability changes when we impose a
specific type of interaction between species.  We start with
predator-prey matrices, which are like random matrices but with the
constraint that if $M_{ij}>0$, then $M_{ji}<0$: the interaction is
beneficial for one species and detrimental for the other. Numerical
simulations showed that these matrices are more stable than random
ones \citep{allesina2008network}. This is confirmed by our results, as
we find that the stability criterion becomes $K<\pi/(\pi-2)\approx
2.75$. Thus, stable predator-prey systems can be much larger and more
complex than random ones. For example, for $\sigma=0.5$, $C=0.1$, the
stability criterion is violated for $S=40$ in the random case, but
$S=304$ for the predator-prey case.

It should be noted that the mean of the off-diagonal coefficients
$\overline{M_{ij}}$ is $0$ in both cases. In fact, both random and
predator-prey matrices will have (on average) the same number of
positive and negative coefficients, and with the same magnitude. The
only difference between the matrices is that in the predator-prey
case, coefficients are arranged in pairs, such that one is negative
and the other is positive. However, this arrangement modifies the
expected interaction strength product for two interacting species
$\overline{M_{ij}M_{ji}}$ (i.e., the expectation is taken over all the
pairs in which $M_{ij}$ or $M_{ji}$ are $\neq 0$), which is $0$ in the
random case, but $-\sigma^22/\pi$ in the predator-prey case. Thus, the
difference in stability arises from having negative
$\overline{M_{ij}M_{ji}}$. This is confirmed by showing that the mean
eigenvalue $\overline{\lambda}=-1$ in both cases (the trace being
$-S$), while the variance \citep{jorgensen2000variance} (for large $S$)
is $\text{Var}(\lambda)=0$ in the random case and
$\text{Var}(\lambda)=(S-1)C\overline{M_{ij}M_{ji}}$ in the
predator-prey case.  Note that the variance can be negative given that
the eigenvalues can be complex conjugate. Having negative variance
means that the variance of the imaginary part of the eigenvalues is
larger than that of the real part. If the stability is driven by
having negative $\overline{M_{ij}M_{ji}}$, reversing its sign should
decrease stability.

The matrices yielding the opposite sign for $\overline{M_{ij}M_{ji}}$,
compared to the predator-prey case, are those in which pairs of
species interact as mutualists or competitors, and for each pair the
interaction type is assigned at random. In these matrices, we still
have $\overline{M_{ij}}=0$, but now $\overline{M_{ij}M_{ji}}=\sigma^2
2/\pi$. Accordingly, the stability criterion becomes
$K<\pi/(\pi+2)\approx 0.61$: this mixture of competition and mutualism
leads to a large decrease in stability. For $\sigma=0.5$, $C=0.1$ the
criterion is violated for $S=15$.

We derived these criteria for stability in the following way. Consider
a random, $S\times S$ matrix, $A$, whose elements are all Gaussian
with mean $\overline{A_{ij}}=0$, mean square value
$\overline{A_{ij}^2}=1/S$, and mean interaction strength product
$\overline{A_{ij}A_{ji}}=\tau /S$.  For $S \rightarrow \infty$, the
eigenvalues of $A$, $\lambda=x+iy$, are uniformly distributed in the
ellipse $(x/a)^2+(y/b)^2 \leq 1$, with $a=(1+\tau)$ and $b=(1-\tau)$
\citep{sommers1988spectrum}.

To obtain the community matrices we are interested in, rescale the
matrix $A$: $\sigma \sqrt{S}A=M$. Thus, the elements $M_{ij}$ have the
following properties: $\overline{M_{ij}}=0$,
$\overline{M_{ij}^2}=\sigma^2$, and $\overline{M_{ij}M_{ji}}=\tau
\sigma^2$. For large $S$, the eigenvalues of $M$ are approximately
uniformly distributed in an ellipse with $a=\sigma \sqrt{S} (1+\tau)$
and $b= \sigma \sqrt{S} (1-\tau)$.

The value of $\tau$ can be derived for all the types of matrices
illustrated above. In the random case, $\overline{M_{ij}M_{ji}}=\tau
\sigma^2=0$, and thus $a=b=\sigma \sqrt{S}$ (i.e., the eigenvalues are
distributed in a circle). For the predator-prey case, the expectation
for the product $\overline{M_{ij}M_{ji}}$ is $-\sigma^2 2/\pi$: the
expectation for the product of two independent, identically
distributed half-normal random variables, with a negative sign
accounting for the opposite signs of the coefficients. Thus,
$\overline{M_{ij}M_{ji}}=\tau \sigma^2=-\sigma^2 2/\pi$, leading to
$\tau =- 2/\pi$, $a=\sigma\sqrt{S}(1-2/\pi)$ and
$b=\sigma\sqrt{S}(1+2/\pi)$. Similarly, for the mixture of competition
and mutualism we have $\overline{M_{ij}M_{ji}}=\tau \sigma^2=\sigma^2
2/\pi$ and thus $\tau =2/\pi$.

These results hold for completely connected matrices, with ellipses
centered at $(0,0)$. Setting the diagonal coefficients to $-d$ centers
the ellipses at $(-d,0)$ (for May's results, $d=1$). For stable
matrices, the ellipses must be fully contained in the left half-plane
($a < d$), as the real part of all eigenvalues must be negative to
attain stability. Accordingly, the stability criteria for fully
connected matrices are: random, $\sigma\sqrt{S}<d$; predator-prey,
$\sigma\sqrt{S} (1-2/\pi)<d$; mixture of competition and mutualism,
$\sigma\sqrt{S}(1+2/\pi)<d$. To account for general connectance,
$C<1$, we follow May \citep{may1972will,may2001stability} and include
it under the square root, obtaining the stability criteria in Table 1.

\begin{table}
{\footnotesize
  \begin{tabular}{rc|ccc}

    $\,$ & $\,$ & \multicolumn{3}{c}{$S_{\text{max}}$} \\
    Interaction & Stability Criterion & 
    {\footnotesize $(C,\sigma,d)=(0.33,0.1,0.25)$}  
    & {\footnotesize $(0.25,0.2,1.0)$} 
    & {\footnotesize $(0.25,0.5,2.0)$}\\ \hline 
    Nested Mut. & & 8 & 20 & 16 \\
    Mutualism & $\sigma (S-1)C<\frac{d}{\sqrt{\frac{2}{\pi}}}$ & 
    8 (8.5) & 23 (24) & 18 (19.05) \\
    Bipartite Mut. & & 9 & 24 & 19 \\
    Mixture & $\sigma \sqrt{SC}<\frac{d\pi}{\pi+2}$ & 
    11 (7) & 28 (23.9) & 41 (37.3)\\
    Competition & $\sigma \left( \frac{\sqrt{S}\pi}{2+\pi}+\sqrt{\frac{2}{\pi}} \right)<d$ 
    & 12 & 33 & 50\\
    Random & $\sigma \sqrt{SC}<d$ & 
    26 (18.9) & 72 (64) & 109 (100)\\
    Niche Pred.-Prey & & 
    75 & 202 & 295 \\
    Cascade Pred.-Prey & & 
    125 & 417 & 649 \\
    Predator-Prey & $\sigma \sqrt{SC}<\frac{d\pi}{\pi-2}$ & 
    148 (143.4) & 482 (484.7) & 745 (757.3)\\
    \hline
  \end{tabular}
}
  \caption{{\small \em Stability criteria for different types of interactions and
      network structures. In all cases, the criteria hold for large,
    $S \times S$ Gaussian matrices with connectance $C$ and diagonal
    coefficients $-d$. The elements of the matrix have mean square
    $\sigma^2$. The competition criterion holds for $C=1$. Numerical
    simulations report, for a given combination of parameters, the
    largest $S$ yielding probability of stability $\geq 0.5$ (computed
    using 1000 matrices). In parenthesis the analytical predictions.}}
\end{table}

We have confirmed these results by plotting the density of the
eigenvalues in the complex plane (Figure 1, top): even for matrices of
moderate size (50 species or more), the approximation is very
accurate. To show the sharpness of the transition from high to low
probability of stability, we performed extensive numerical simulations
(Figure 1, bottom).

\begin{figure}
  \includegraphics[width=\linewidth]{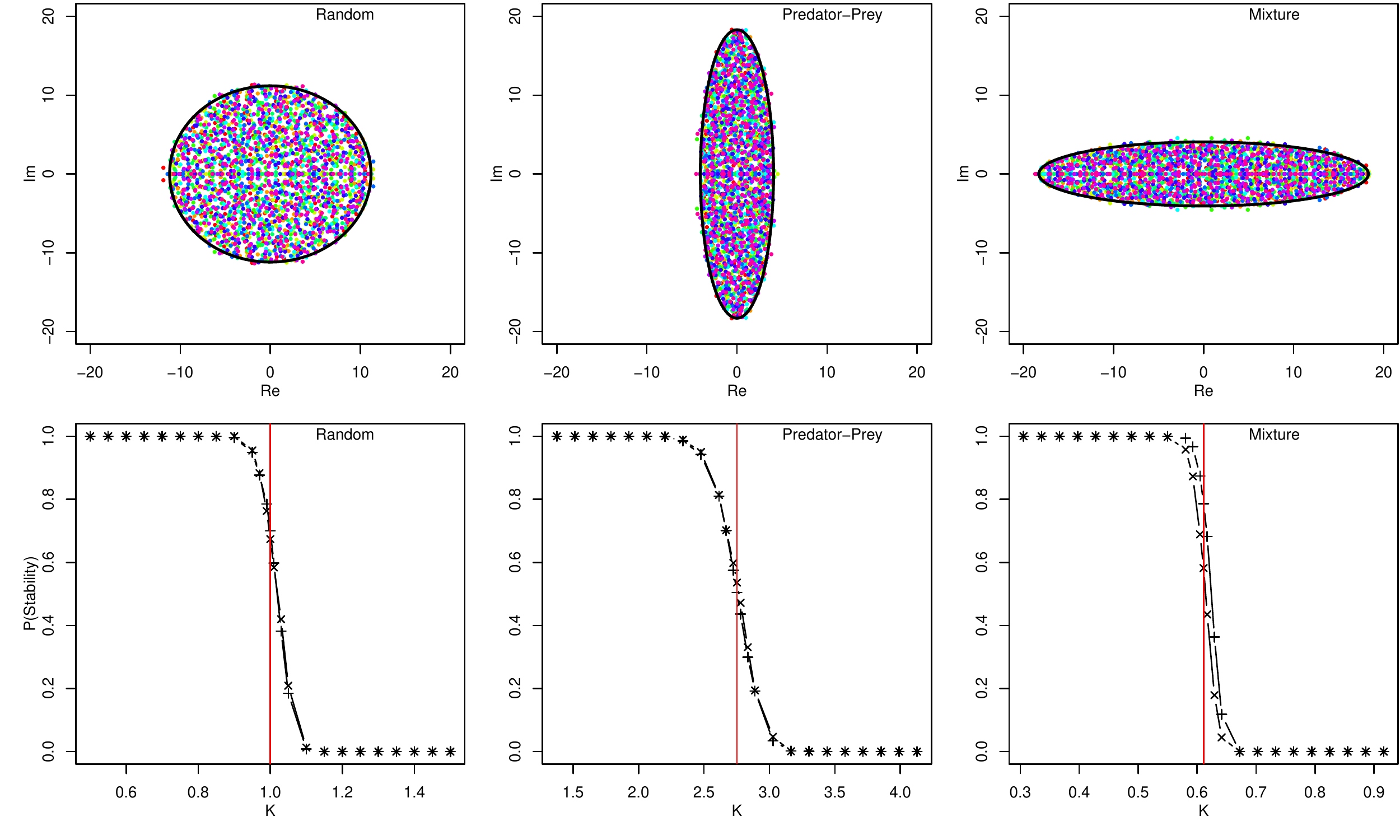}
  \caption{{\small \em Top: Distribution of the eigenvalues for random,
    predator-prey and mixture of competition and mutualism
    matrices. For $S=250$, $C=0.25$ and $\sigma=1$, we plot the
    eigenvalues of 10 matrices (colors) with $0$ on the diagonal and
    off-diagonal elements following the random, predator prey or
    mixture prescriptions. The black ellipses are derived analytically
    in the text. Bottom: Corresponding stability profiles. For the
    random case, starting from $S=250$, $C=0.5$, $\sigma=0.1$ and
    $d=1$, we systematically varied $C$ ($\times$) or $\sigma$ ($+$)
    in order obtain $K=\sigma \sqrt{SC}$ spanning $[0.5,\ldots, 1.0,
      \ldots, 1.5]$ of the critical value for stability (indicated in
    red, 1 in the case of random matrices). The profiles were obtained
    computing the probability of stability out of 1000 matrices. The
    predator-prey case is as the random but with $\sigma=0.5$ and
    critical value $\pi/(\pi-2)$. The mixture case is as the random
    but with critical value $\pi/(\pi+2)$.}}
\end{figure}

In all the above cases, $\overline{M_{ij}}=0$, and the expected row
(column) sum is also $0$. What happens if we relax these constraints?
The most extreme case is that of mutualism ($M_{ii}=-d$, whereas
$M_{ji},M_{ij}$ are drawn from $|N(0,\sigma^2)|$ with probability $C$
and zero otherwise), in which the mean coefficient is
$\overline{M_{ij}}=\sigma \sqrt{2/\pi}$ and the expected row (column)
sum is $R=(S-1)C \overline{M_{ij}}-d$. In these matrices, we find an
extreme, real eigenvalue $\lambda^P=R$, while the remaining
eigenvalues (for $C=1$) are arranged in a circle centered at
$(-d-C\overline{M_{ij}},0)$ (Figure 2, top). Numerical simulations
suggest that the radius of this circle is approximately $K
\pi/(2+\pi)$. Because for mutualism the stability is determined
exclusively by $\lambda^P=R$, the criterion becomes $R<0$, which is
equivalent to diagonal dominance \citep{vargascircle}. Therefore, in
this type of matrix the fact that interactions are arranged in pairs
does not influence stability. For the competition case, the situation
is reversed (Figure 2, bottom): we find an extreme negative eigenvalue
$\lambda^N = R =-(S-1)C \overline{M_{ij}}-d$, and the others are
contained (for $C=1$), in a circle centered in
$(-d+C\overline{M_{ij}},0)$ with a radius of $K \pi/(2+\pi)$. The
maximum eigenvalue is at the very right edge of the circle, so that we
can derive a stability criterion only for $C=1$:
$\sigma(\sqrt{S}\pi/(2+\pi)+\sqrt{2/\pi})<d$. For the general case in
which $C< 1$, the non-extreme eigenvalues are approximately contained
in an ellipse, but we have not found an exact expression for general
$C$.

\begin{figure}
  \includegraphics[width=\linewidth]{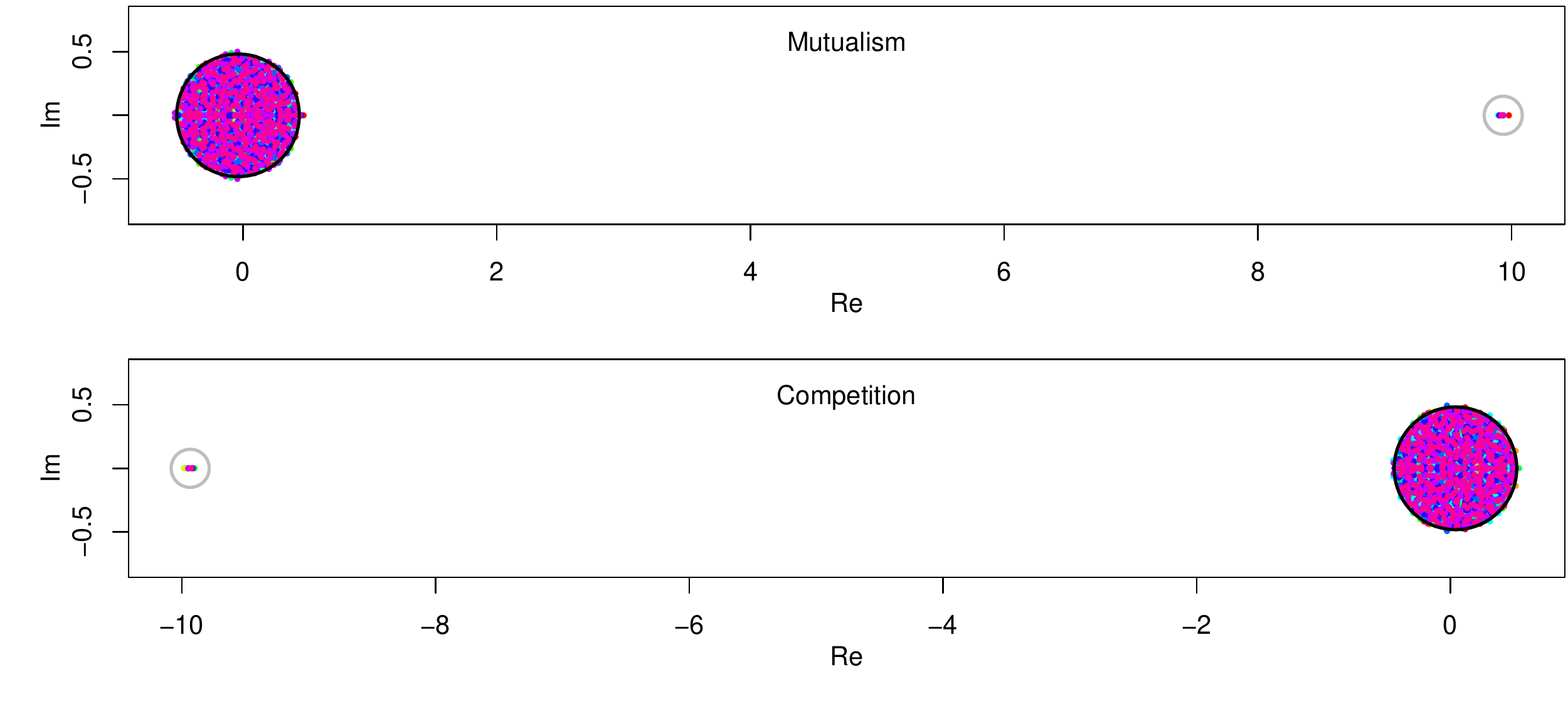}
\caption{{\small \em Distribution of the eigenvalues for mutualism and
  competition, obtained from 10 matrices (colors), with $S=250$,
  $C=1$, $\sigma=0.05$ and $d=0$. In both cases we observe an extreme,
  real eigenvalue whose value equals the row-sum in the matrices
  (circled in gray). The remaining eigenvalues, for $C=1$, are
  contained in a circle of radius $\sigma \sqrt{SC} \pi / (2+\pi)$.}}
\end{figure}

Returning to the predator-prey case, we assess the effect of imposing
realistic food web structure. In community matrices describing food
webs, we expect producers to have positive columns and negative rows,
with the converse for top predators. These variations are likely to
move some eigenvalues ``vertically'' in the complex plane (as the
large row and column sums have opposite signs). To test this
hypothesis, we plotted the eigenvalues for predator-prey webs in which
interactions are arranged following the cascade
\citep{cohen1990community} and niche \citep{williams2000simple} models
(Appendix). Although these models cannot fully reproduce food web
structure
\citep{williams2000simple,cattin2004phylogenetic,allesina2008general},
they are widely used to simulate food webs (e.g.,
\citep{chen2001transient,martinez2006diversity}).  It is generally
believed that including realistic food web structure should increase
stability (e.g., \citep{mcnaughton1978stability,yodzis1981stability}).

In both cases, we find several extreme eigenvalues with large
imaginary part (Figure 3). Because the eigenvalues of these matrices
must have the same mean and variance \citep{jorgensen2000variance}
observed in the unstructured predator-prey case, we observe a
distortion of the ellipses, yielding larger real parts than
expected. Thus, cascade and niche models should produce networks that
are less stable than their unstructured predator-prey counterpart,
with the niche model having a larger discrepancy. These considerations
are confirmed by numerical experiments (Figure 3).  We conclude that,
surprisingly, imposing realistic food web structure hampers stability.

In the same spirit, we measured the effect realistic structures
produce in mutualistic networks. Several published mutualistic
networks are bipartite
\citep{bascompte2003nested,okuyama2008network,bastolla2009architecture,thebault2010stability}:
there are two types of nodes (e.g., plants and pollinators), and
interactions occur exclusively between different types. Also,
bipartite mutualistic networks tend to be nested
\citep{bascompte2003nested}: the interactions of the specialists are
subset of those of the generalists. Nestedness is believed to beget
stability
\citep{okuyama2008network,bastolla2009architecture,thebault2010stability}. We
drew the eigenvalues for these two types of structure (bipartite,
bipartite and nested Appendix, Figure 4), and contrasted the results
with those obtained for the unstructured mutualistic case. We observed
several large real eigenvalues as in the unstructured mutualistic
case, but now for each positive eigenvalue we found an equally large
negative one (Figure 4). The bipartite case yields row sums that are
substantially similar to the unstructured case. Accordingly, we do not
expect a large discrepancy in stability. However, given that nested
structures will yield (on average), larger maximum row/column sum
(associated with the generalist plants and animals), nested structures
are inherently less stable than unstructured ones.  These results are
confirmed by numerical simulations (Figure 5).

We have so far considered how the arrangement of the interaction
coefficients affects stability, we now assess the role of interaction
strength distributions. We have extracted coefficients from normal (or
half-normal) distributions, where the majority of interactions are
close to 0 and thus ``weak''. These ``weak interactions'' are thought
to contribute considerably to the stability and persistence of natural
systems \citep{mccann1998weak,emmerson2004weak}. To examine their
effect, we extended our analysis to the cases in which the absolute
value of each coefficient is taken either from a uniform or a gamma
distribution, parametrized such that $\overline{M_{ij}}=0$ and
$\overline{M_{ij}^2}=\sigma^2$ (as in the normal case)
(Appendix). Note that the different distributions represent different
frequencies of weak interactions. Therefore, changing the distribution
impacts $\overline{M_{ij}M_{ji}}$, which, compared to the random case,
has to be negative to increase stability and positive to depress
stability. When weak interactions are preponderant,
$\overline{M_{ij}M_{ji}}$ is expected to be smaller in magnitude.
Therefore, weak interactions should increase the stability for
mutualistic and competitive systems, but decrease the stability of
predator-prey matrices and have no effect in the random case. This
argument is matched by analytical and numerical results (Appendix),
showing that, contrary to the current belief, weak interactions can be
destabilizing.

To summarize the stability properties of the various matrices, we
performed numerical simulations for all the types of networks and with
three parametrizations (Table 1). We searched for the largest $S$
yielding a probability of stability $\geq 0.5$ (measured using 1000
matrices). In all cases, $S$ increases when moving from nested
mutualism to predator-prey.

We have shown that arrangement into signed pairs of interactions has
large impact on stability. For example, the random, mixture and
predator-prey matrices contain basically the same coefficients, and
the large difference in stability is driven exclusively by their
arrangement (pairs with random signs, pairs with same sign and pairs
with opposite signs, respectively). This is consistent with the fact
that the variance of the eigenvalues is driven by the mean product
between nonzero pairs $\overline{M_{ij}M_{ji}}$.

Mutualism, competition, and their mixture, although yielding the same
pairwise mean product, and the same mean and variance of the
eigenvalues, have very different stability properties: increasing the
fraction of competitive interactions increases stability. In these
cases the stability is driven by higher moments of the eigenvalue
distribution. We conjecture that the stability properties in these
matrices are thus influenced by products of three or more coefficients
at a time.  For example, the product of three competitive interactions
would yield a different sign from that of three mutualistic
interactions, potentially accounting for the difference in stability.

Surprisingly, imposing realistic structure to the interaction networks
appears to be detrimental for stability both in the predator-prey and
the mutualistic cases. This does necessarily mean that more realistic
networks should be less stable, as in our comparisons all the
coefficients have (in absolute value) the same expectation, while this
is not the case in natural systems (e.g., generalist species will have
typically lower values for each interaction compared to
specialists). However, we can safely assert that realistic structure
alone is not contributing to stability: an increase in stability can
be observed only if there is an interplay between the network
structure and the interaction strengths. The fact that interaction
strengths are a major determinant of stability is confirmed by our
analysis showing that weak interactions can be either stabilizing or
destabilizing depending on the type of interaction between species.

We have shown that simply plotting the density of the eigenvalues
provides qualitative insight into the stability of the systems. Using
this graphical method, the effect of any type of network structure can
be readily analyzed.

Finally, we have found a consistent ``stability hierarchy'' spanning
mutualism to predator-prey. Predator-prey interactions enable the
stable coexistence of networks as large and complex as those observed
empirically.

Our results are not limited to the stability of ecological (or
biological) systems. The criteria, in fact, hold for any system of
differential equations resting at an equilibrium point.

\bibliographystyle{elsarticle-num}

\section*{Appendix}
\noindent{\bf Construction of the community matrices}

In the main text we analyze different types of matrices. Here we
detail how the matrices were constructed. In all cases, the parameters
are: $S$, number of species; $C$, desired level of connectance;
$\sigma$, standard deviation of the normal distribution from which
coefficients are drawn; $-d$ value of the diagonal coefficients. For
each matrix, we also report the expected mean and variance for the
eigenvalues.

\noindent{\em Random Matrices}

\noindent In the random case, we construct the matrices in the
following way: i) for each coefficient $M_{ij}$, $i \neq j$, we draw a
random value $X$ from an uniform distribution $U[0,1]$. If the value
is $X\leq C$, we draw the coefficient $M_{ij}$ from a normal
$N(0,\sigma^2)$. Otherwise ($X>C$), $M_{ij}=0$. ii) All
$M_{ii}=-d$. For these matrices, the eigenvalues have
$\overline{\lambda}=-d$ and $\text{Var}(\lambda)=0$.

\noindent{\em Predator-Prey Matrices}

\noindent i) For each coefficient $M_{ij}$, $j>i$, we draw a random
value $X$ from $U[0,1]$. ii) If the value is $X \leq C$, we draw a
second random value $Y$ from $U[0,1]$. If this new random value is $Y
\leq 0.5$, we draw $M_{ij}$ from an half-normal distribution
$|N(0,\sigma^2)|$ and $M_{ji}$ from a negative half-normal
$-|N(0,\sigma^2)|$, while if $Y>0.5$ we do the opposite. iii) If
$X>C$, we assign $0$ to both coefficients. iv) All $M_{ii}=-d$. For
predator-prey, $\overline{\lambda}=-d$ and
$\text{Var}(\lambda)=-(S-1)C\sigma^22/\pi$.

\noindent{\em Mixture of Competition and Mutualism Matrices}

\noindent i) For each coefficient $M_{ij}$, $j>i$, we draw a random
value $X$ from $U[0,1]$. ii) If the value is $X \leq C$, we draw a
second random value $Y$ from $U[0,1]$. If this new random value is $Y
\leq 0.5$, we draw $M_{ij}$ and $M_{ji}$ from an half-normal
$|N(0,\sigma^2)|$, while if $Y>0.5$ we draw both coefficients from a
negative half-normal distribution. iii) If $X>C$, we assign $0$ to
both coefficients. iv) All $M_{ii}=-d$. For this mixture,
$\overline{\lambda}=-d$ and $\text{Var}(\lambda)=(S-1)C\sigma^22/\pi$.

\noindent{\em Mutualism Matrices}

\noindent i) For each coefficient $M_{ij}$, $j>i$, we draw a random
value $X$ from $U[0,1]$. If the value is $X \leq C$, we draw both
$M_{ij}$ and $M_{ji}$ from an half-normal
$|N(0,\sigma^2)|$. Otherwise, we assign $0$ to both coefficients. ii)
All $M_{ii}=-d$. For mutualism, $\overline{\lambda}=-d$ and
$\text{Var}(\lambda)=(S-1)C\sigma^22/\pi$.

\noindent{\em Competition Matrices}

\noindent i) For each coefficient $M_{ij}$, $j>i$, we draw a random
value $X$ from $U[0,1]$. If the value is $X \leq C$, we draw both
$M_{ij}$ and $M_{ji}$ from a negative half-normal
$-|N(0,\sigma^2)|$. Otherwise, we assign $0$ to both coefficients. ii)
All $M_{ii}=-d$. For competition, $\overline{\lambda}=-d$ and
$\text{Var}(\lambda)=(S-1)C\sigma^22/\pi$.

\noindent{\em Cascade Predator-Prey Matrices}

\noindent In the cascade model \citep{cohen1990community}, species are
ordered and each species has a fixed probability of preying upon the
preceding species.  The produced networks do not contain cycles,
although cycles are observed in empirical networks
\citep{williams2000simple}. In the cascade model, the species with
highest ranking functions as a top predator, while that with the
lowest ranking as a producer. Accordingly, the highest ranked has
positive column and negative row, while the opposite is true for the
lowest ranked.

The matrix construction algorithm is: i) For each coefficient
$M_{ij}$, $j>i$, we draw a random value $X$ from $U[0,1]$. If the
value is $X \leq C$, we draw the coefficient $M_{ij}$ from an
half-normal $|N(0,\sigma^2)|$ and the coefficient $M_{ji}$ from a
negative half-normal $-|N(0,\sigma^2)|$. Otherwise, we assign $0$ to
both coefficients. ii) All $M_{ii}=-d$. For the cascade model,
$\overline{\lambda}=-d$ and
$\text{Var}(\lambda)=-(S-1)C\sigma^22/\pi$.

\noindent{\em Niche Predator-Prey Matrices}

\noindent The niche model \citep{williams2000simple} allows for trophic
cycles and cannibalism. The species are ordered (each one being
assigned a ``niche value'', $\eta_i$). A ``niche radius'', $r_i$,
proportional to $\eta_i$, is drawn for each species along with a
``niche center'' $c_i$. Each species $i$ preys upon all the species
whose $\eta_j$ are included in the range $[c_i-r_i/2,c_i+r_i/2]$. The
produced networks are interval (i.e., each predator preys upon
consecutive species). Empirical networks, however, are not perfectly
interval
\citep{williams2000simple,cattin2004phylogenetic,allesina2008general}.

To generate the matrices, we first produced an adjacency matrix $A$,
using the niche model ($A_{ij}=1$ if $i$ is a prey of $j$). Then we
obtained a ``sign matrix'' $S=-A+A^t$. Finally, $M_{ij}$ is obtained
multiplying $X_{ij}$ taken from an half-normal distribution
$|N(0,\sigma^2)|$ by $S_{ij}$. The diagonal elements are set to $-d$.
For the niche model, $\overline{\lambda}=-d$ and
$\text{Var}(\lambda)=-(S-1)C\sigma^22/\pi$.

\noindent{\em Mutualistic Bipartite Matrices}

\noindent For the bipartite case, we divided the species in two group
of equal size ($S/2$, when $S$ is even). For each $M_{ij}$ where $i$
belongs to the first group and $j$ to the second, we draw $M_{ij}$ and
$M_{ji}$ from an half-normal distribution $|N(0,\sigma^2)|$ with
probability $C'=2C(S-1)/S$ (so that the expected connectance is
matched). The diagonal is $-d$. For the bipartite mutualistic model,
$\overline{\lambda}=-d$ and $\text{Var}(\lambda)=(S-1)C\sigma^22/\pi$.

\noindent{\em Mutualistic Nested Matrices}

\noindent Nestedness is a property of the incidence matrix $B$
(typically rectangular) in which the row are the species belonging to
the first group (e.g., plants) and the rows those in the second group
(e.g., pollinators). Say that to match the desired connectance $C$ in
the matrix $M$, we want to arrange $L$ links in $B$. We arrange the
links in the following way: First, to guarantee connectedness, i) we
fill the first row; ii) we fill the first column; iii) We arrange the
subsequent links so that the matrix is perfectly nested. For example
(using a squared incidence matrix), say that $B$ is $6 \times 6$:
\begin{equation*}
  B=\left[
  \begin{tabular}{cccccc}
    {\bf 1,1} & {\bf 1,2} & {\bf 1,3} & {\bf 1,4} & {\bf 1,5} & {\bf 1,6}\\
    {\bf 2,1} & {\bf 2,2} & {\bf 2,3} & {\bf 2,4} & {\bf 2,5} & 2,6\\
    {\bf 3,1} & {\bf 3,2} & {\bf 3,3} & 3,4 & 3,5 & 3,6\\
    {\bf 4,1} & {\bf 4,2} & 4,3 & 4,4 & 4,5 & 4,6\\
    {\bf 5,1} & 5,2 & 5,3 & 5,4 & 5,5 & 5,6\\
    {\bf 6,1} & 6,2 & 6,3 & 6,4 & 6,5 & 6,6\\
  \end{tabular}
  \right]
\end{equation*}
and suppose we want to include 17 links. First, we fill the first row
$(1,1)$ to $(1,6)$, so that we placed 6 links. The next five links are
used to fill the first column $(2,1)$ to $(6,1)$. Finally, the last
six links are placed in $(2,2)$, $(2,3)$, $(3,2)$, $(2,4)$, $(3,3)$,
etc. Note that the sum of the $x$ and $y$ coordinates for the links is
growing. In fact, ordering the potential link by their coordinate
sums, and giving precedence to those with smaller row number in case
of ties, guarantees the maintenance of perfect nestedness. This is the
filling algorithm we used in the simulations. Once we obtain $B$, we
use it (along with its transpose), to determine the interactions in
the matrix $M$. All the nonzero values of $M_{ij}$ are taken from the
half-normal $|N(0,\sigma^2)|$. The diagonal is $-d$. For the bipartite
nested mutualistic model, $\overline{\lambda}=-d$ and
$\text{Var}(\lambda)=(S-1)C\sigma^22/\pi$.

\noindent{\bf Different Distributions and Weak Interactions} 

The goal of this section is to investigate whether our findings are
robust to changes in the distribution of interaction strengths and to
assess the role weak interactions play for stability. In the main
text, we deal with normal (or half-normal) distributions. Here, we
consider uniform distributions and gamma distributions with different
shapes.

To measure how preponderant ``weak interactions'' are in a
distribution of interaction strengths, we take the expectation for the
absolute value. For example, for a normal distribution, define $X$ as
a random variable taken from the distribution $|N(0,\sigma^2)|$. The
expectation for $X$ is $E[X]=\sigma \sqrt{2/\pi} \approx 0.798 \sigma$
(as we saw in the main text).

Before we can examine the other distributions, we have to parametrize
them in such a way that $\overline{M_{ij}}=0$ and
$\overline{M_{ij}^2}=\sigma^2$. In this way, we are satisfying the
conditions we stated in the main text, and we are considering
distributions of interaction strengths with the same mean and
variance.

We start from a uniform distribution $U[0,\theta]$. We will sample
positive coefficients from the distribution and negative coefficients
reversing the sign. Thus, the interaction strengths (for the nonzero
terms) are distributed uniformly in $U[-\theta,\theta]$. Clearly, the
distribution satisfies $\overline{M_{ij}}=0$. When
$\theta=\sigma\sqrt{3}$, the variance $\overline{M_{ij}^2}=
\frac{4}{12}\theta^2\sigma^2=\sigma^2$. With such a parametrization,
the expected interaction is $E[X]=\sigma\sqrt{3}/2 \approx 0.886
\sigma$: typically, interactions will be larger than in the normal
case. What are the consequences for stability? We begin by computing
$\tau$. For a predator-prey matrix whose elements are taken from the
uniform distribution above, the expected product of the interaction
strengths of two interacting species is
$\overline{M_{ij}M_{ji}}=-\frac{3}{4} \sigma^2$ and thus
$\tau=-\frac{3}{4}$. This means that the stability criterion for
uniform predator-prey becomes $\sigma\sqrt{SC}<4d$: the uniform
predator-prey matrices are much more likely to be stable than their
normal counterpart. The uniform random and mixture cases trivially
follow. For the uniform random, $\sigma\sqrt{SC}<d$: exactly as in the
normal case. For the uniform mixture, $\sigma\sqrt{SC}<4d/7$, less
than in the normal case. To summarize, for the uniform distribution
(where we expect stronger interactions than in the normal case), we
observe an increase in stability for the predator-prey case and a
decrease for the mixture case. The random case in unaltered.

To further prove that weak interactions stabilize the mixture and
destabilize the predator-prey cases, we analyzed the effect of taking
the magnitude of the coefficients from a gamma distribution. The gamma
distribution takes two parameters, $k$ (shape), and $\theta$
(scale). As above, we want to ensure that $\overline{M_{ij}}=0$ and
$\overline{M_{ij}^2}= \sigma^2$. This is accomplished, for arbitrary
${k}$, by choosing $\theta=\sigma \sqrt{\frac{1}{k(k+1)}}$. For such a
gamma distribution, $E[X]=\sigma \sqrt{\frac{k}{k+1}}$. For example,
for $k=0.7$, $E[X]\approx 0.642 \sigma$ (less than in the normal
case), while for $k=3$, $E[X]\approx 0.866 \sigma$ (more than in the
normal case, and exactly as in the uniform case). Computing $\tau$ for
the predator-prey case, we find $\tau=-k/(k+1)$ and thus the stability
criterion becomes $\sigma\sqrt{SC}<(k+1)d$. This means a higher
likelihood of stability, compared to the normal case, whenever
$k+1>\pi/(\pi-2)$ (approximately, $k>1.75$). Thus, when we extract the
magnitude of the coefficients from a gamma distribution, increasing
$k$ will increase the expectation, and this in turn will result in
more stability for the predator-prey case and less for the mixture
(criterion: $\sigma \sqrt{SC}<d(2k+1)/(k+1)$). For the random case the
situation is unaltered.

Our analytical predictions are confirmed by plotting the density of
the eigenvalues (Figure 6) and drawing stability profiles (Figure 7)
for all the distributions described above.

\begin{figure}
\includegraphics[width=\linewidth]{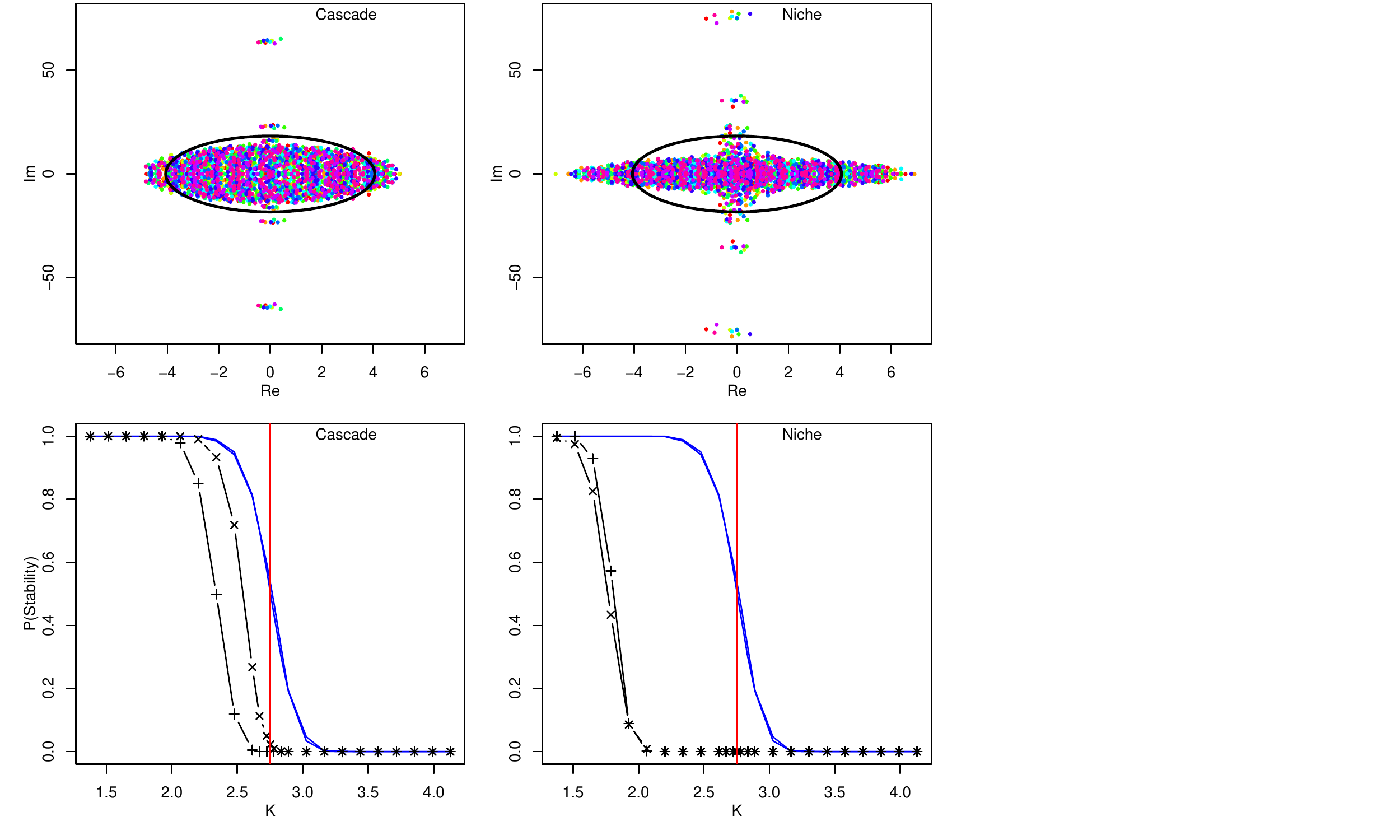}
\caption{{\small \em Top: Distribution of the eigenvalues for cascade
    and niche models, with the same values used in Figure 1. In both
    cases we observe extreme, largely imaginary
    eigenvalues. Accordingly, the theoretical ellipse for the
    predator-prey case (black) does not contain all the remaining
    eigenvalues, decreasing the probability of stability. This effect
    is more pronounced in the niche case. Bottom: Stability profiles
    obtained using the same values as Figure 1 (unstructured
    predator-prey case). We report the unstructured predator-prey
    profile (blue) for comparison. Note that, contrary to the
    unstructured model, in the cascade and niche matrices increasing
    $C$ or $\sigma$ yields different effects (separation between the
    two lines). This is because the row (column) sum scales linearly
    with $C$ and $\sigma$, while $K$ scales sublinearly with $C$.}}
\end{figure}

\begin{figure}
\includegraphics[width=\linewidth]{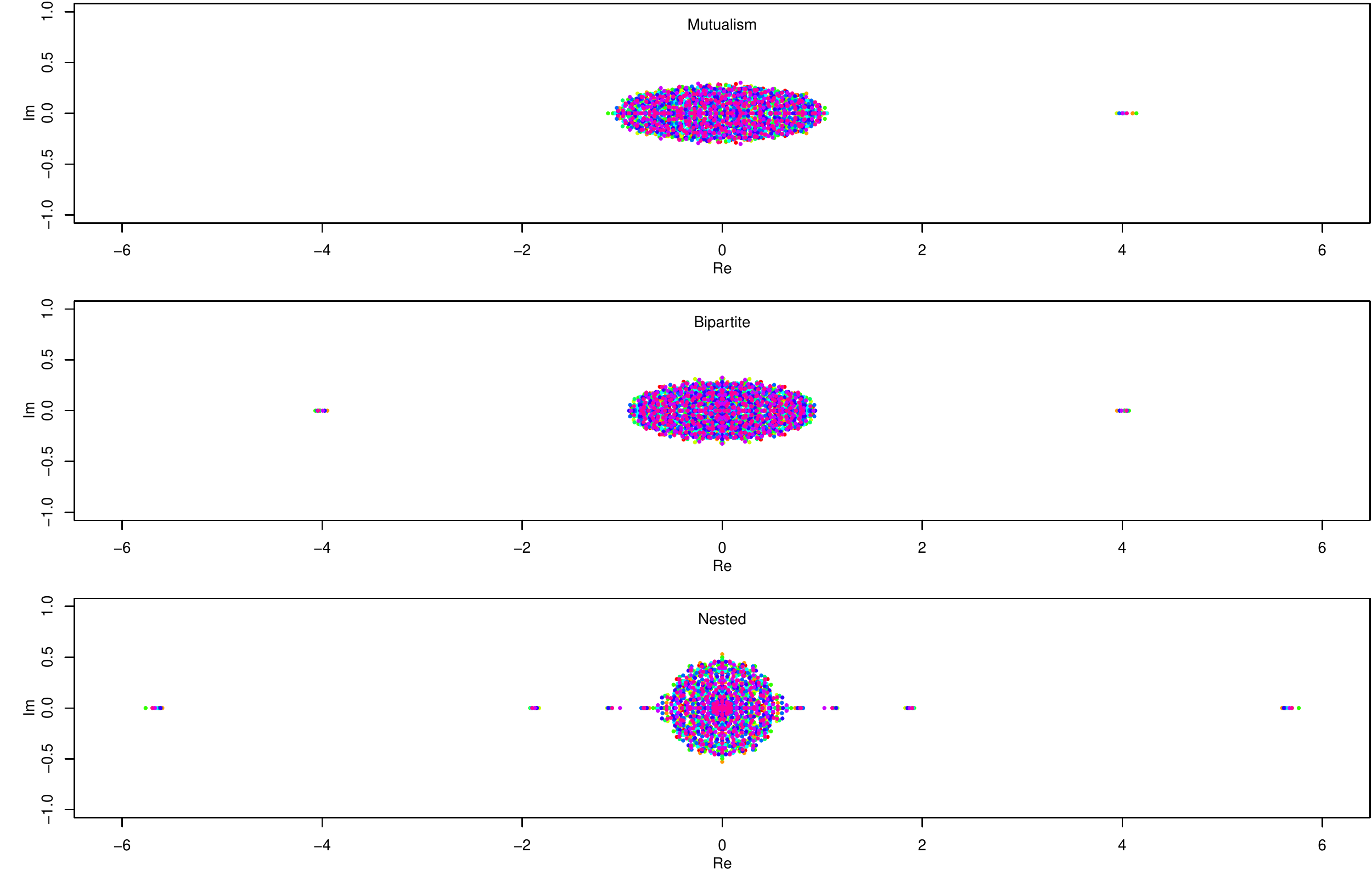}
\caption{{\small \em Distribution of the eigenvalues for the
    unstructured mutualism, bipartite mutualism and nested
    mutualism. In all cases, $S=250$, $\sigma=0.1$, $C=0.2$ and
    $d=0$. Note that the bipartite case does produce extreme negative
    real eigenvalues coupled with positive ones, but the row sum (and
    thus the dominant eigenvalue) is similar to that of the
    unstructured mutualistic case. The nested matrices, in which
    generalist species yield (on average), larger row and column sums,
    display larger dominant eigenvalues. Thus, nestedness should
    produce matrices that are less likely to be stable compared to the
    other two cases.}}
\end{figure}

\begin{figure}
  \includegraphics[width=0.5\linewidth]{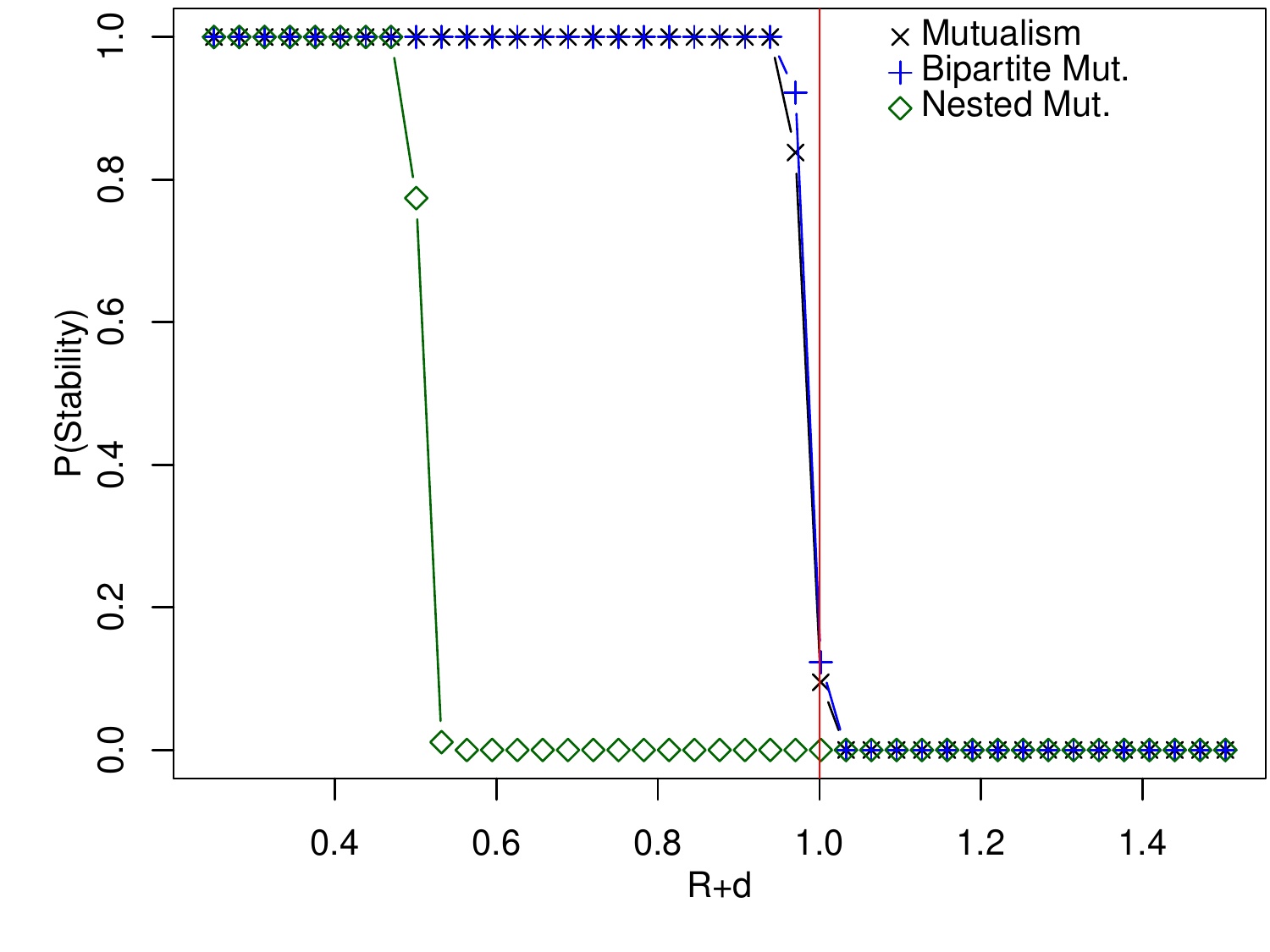}
  \caption{{\small \em Stability profile for the mutualistic cases,
      for $S=250$, $\sigma=0.025$, $d=1$. We vary $C$ so that the
      critical value $(S-1)C\sigma \sqrt{2/\pi}=R+d$ (x-axis) spans
      $[0.2,1.2]$. The critical value is reached for $(S-1)C\sigma
      \sqrt{2/\pi}=1$ (red line). Note that, as expected from Figure
      4, nested matrices are much less stable than the other two types
      of matrices.}}
\end{figure}

\begin{figure}
  \includegraphics[width=0.9\textheight,angle=-90]{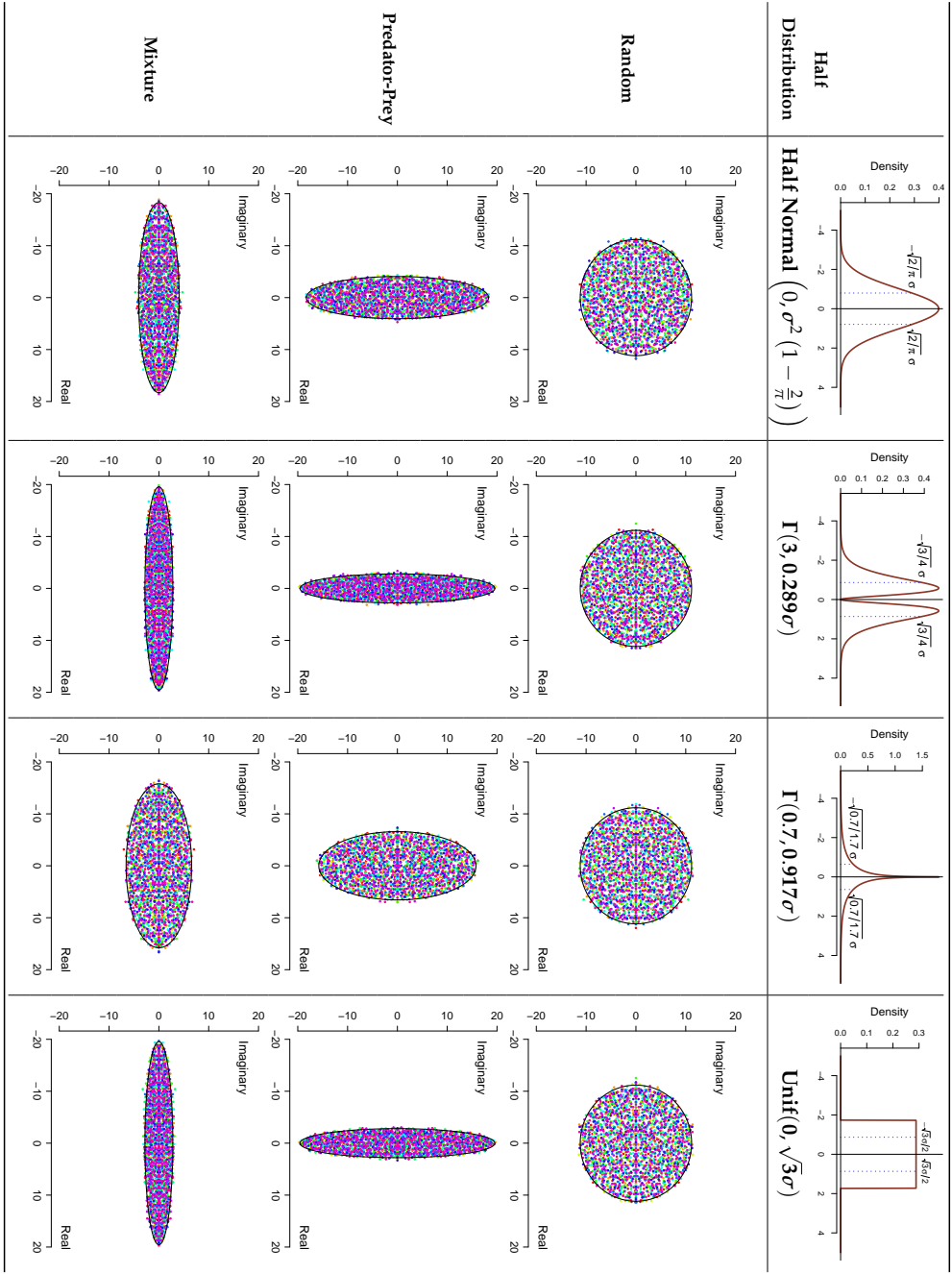}
  \caption{{\small \em Distribution of the eigenvalues for
    random, predator-prey and mixture of competition and mutualism
    matrices (rows) for different distributions (top row). For $S=250$,
    $C=0.25$ and $\sigma=1$, we plot the eigenvalues of 10 matrices
    (colors) with $0$ on the diagonal. For the off-diagonal elements, the
    magnitudes are taken from an half-normal (first column), gamma (second
    and third columns) or uniform (fourth column) distributions, while the
    signs are assigned according to the types of matrices. The
    distributions are parametrized in a way such that
    $\overline{M_{ij}}=0$ and $\overline{M_{ij}^2}=\sigma^2$. The ellipses
    are derived in the Appendix.}}
\end{figure}

\begin{figure}
\includegraphics[width=0.9\textheight,angle=-90]{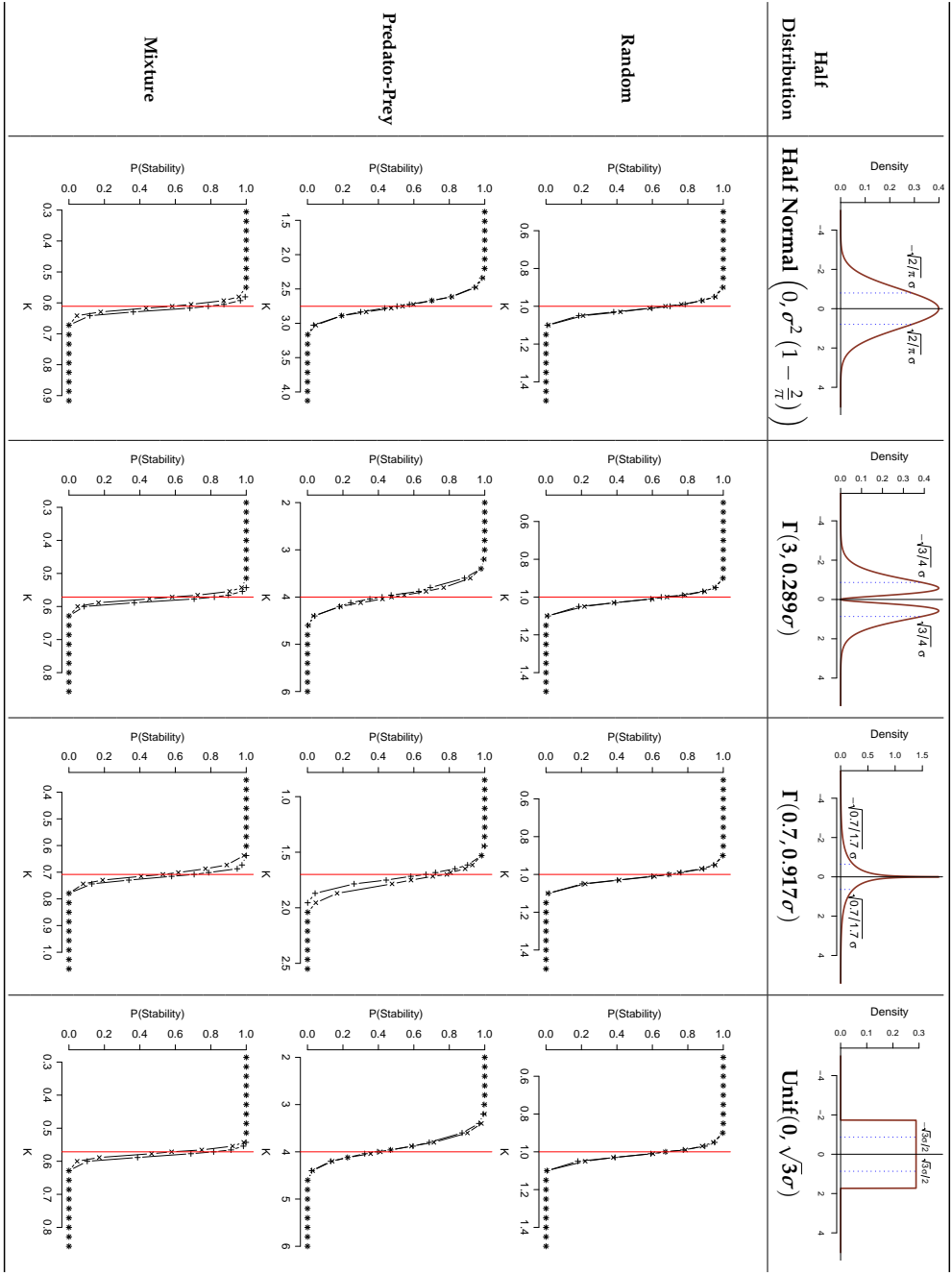}
\caption{Stability profiles for the combinations of matrix type and
  distribution illustrated in Figure 6. For the random and mixture
  cases, starting from $S=250$, $C=0.5$, $\sigma=0.1$ and $d=1$, we
  systematically varied $C$ ($\times$) or $\sigma$ ($+$) in order
  obtain $K=\sigma \sqrt{SC}$ spanning $[0.5,\ldots, 1.0, \ldots,
    1.5]$ of the critical value for stability (indicated in red, 1 in
  the case of random matrices). The profiles were obtained computing
  the probability of stability out of 1000 matrices. The predator-prey
  case is as the random but with $\sigma=0.3$ for the half-normal and
  $\Gamma(0.7,0.917\sigma)$, while $\sigma=0.5$ for the uniform and
  $\Gamma(3,0.289\sigma)$ cases. The adjustment of $\sigma$ is
  necessary as it would otherwise lead to $C>1$.}
\end{figure}

\end{document}